\documentclass[10pt,journal]{IEEEtran} 

\usepackage{textcomp}
\usepackage[utf8]{inputenc}
\usepackage{caption}
\usepackage[perpage,symbol]{footmisc}
\usepackage[english]{babel}
\usepackage{multirow,multicol}
\usepackage{amsmath, bm} 
\usepackage{pbox, booktabs, rotating}
\usepackage{array, makecell, lipsum}
\usepackage{enumerate}
\usepackage{threeparttable}
\usepackage{multirow,hyphenat,balance}
\usepackage{placeins}
\usepackage[compress]{cite}
\usepackage{float,color}
\usepackage{ragged2e}
\usepackage{enumerate}
\usepackage{etoolbox}
\usepackage{paralist}
\usepackage{cases}
\usepackage{url}
\usepackage{framed}
\usepackage{flushend}
\usepackage{amsmath}
\usepackage{amsthm}
\usepackage[english]{babel}
\usepackage{amssymb}
\usepackage{pifont}
\newcommand{\cmark}{\ding{51}}%
\newcommand{\xmark}{\ding{55}}%
\usepackage[linesnumbered,ruled]{algorithm2e}
\let\oldnl\nl
\newcommand{\nonl}{\renewcommand{\nl}{\let\nl\oldnl}}

\theoremstyle{definition}

\newcolumntype{M}[1]{>{\centering\arraybackslash}m{#1}}

\usepackage{commath}
\newcommand*{\ie}{\textit{i.e.,}\@\xspace}
\newcommand*{\eg}{\textit{e.g.,}\@\xspace}
\newcommand*{\el}{et al.\@\xspace}

\DeclareCaptionLabelFormat{nospace}{#1#2}
\author{Yadi Zhong, \IEEEmembership{Student Member IEEE} and Ujjwal Guin, \IEEEmembership{Senior Member IEEE}

\thanks{Yadi Zhong and Ujjwal Guin are with the Department of Electrical and Computer Engineering, Auburn University, AL, USA (e-mail: \{yadi and ujjwal.guin\}@auburn.edu).}

}

\title{Complexity Analysis of the SAT Attack on\\ Logic Locking}

\begin{document}
\maketitle
\thispagestyle{plain}
\pagestyle{plain}

\begin{abstract}
Due to the adoption of horizontal business models following the globalization of semiconductor manufacturing, the overproduction of integrated circuits (ICs) and the piracy of intellectual properties (IPs) can lead to significant damage to the integrity of the semiconductor supply chain. Logic locking emerges as a primary design-for-security measure to counter these threats, where ICs become fully functional only when unlocked with a secret key. However, Boolean satisfiability-based attacks have rendered most locking schemes ineffective. This gives rise to numerous defenses and new locking methods to achieve SAT resiliency. This paper provides a unique perspective on the SAT attack efficiency based on conjunctive normal form (CNF) stored in SAT solver. First, we show how the attack learns new relations between keys in every iteration using distinguishing input patterns and the corresponding oracle responses. The input-output pairs result in new CNF clauses of unknown keys to be appended to the SAT solver, which leads to an exponential reduction in incorrect key values. Second, we demonstrate that the SAT attack can break any locking scheme within linear iteration complexity of key size. Moreover, we show how key constraints on point functions affect the SAT attack complexity. We explain why proper key constraint on AntiSAT reduces the complexity effectively to constant 1. The same constraint helps the breaking of CAS-Lock down to linear iteration complexity. Our analysis provides a new perspective on the capabilities of SAT attack against multiplier benchmark c6288, and we provide new directions to achieve SAT resiliency.
\end{abstract}


\section{Introduction} 
The integrated circuits (ICs) are fundamental to virtually every technology in the Department of Defense (DoD), industrial and commercial spaces. Moore's Law has guided the microelectronics industry for decades to enhance the performance of ICs. The continuous addition of new functionalities in SoCs has forced design houses to adopt newer and lower technology nodes to increase operational speed, reduce power consumption, overall die area, and the resultant cost of a chip. This exponential growth becomes feasible due to the globalization of semiconductor design, manufacturing, and test processes. Building and maintaining a fabrication unit (foundry) requires a multi-billion dollar investment~\cite{ForbesIntel}. As a result, a system-on-a-chip (SoC) design house acquires intellectual properties (IPs) from many vendors and sends the design to a foundry for manufacturing, typically located offshore due to the horizontal integration in the semiconductor industry. At present, the majority of the SoC design houses no longer design the complete SoC and manufacture chips on their own. As a result, the trusted foundry model is no longer assumed to be valid for producing ICs, where the trustworthiness of microelectronic parts is often questioned.

Due to the outsourced IC design and fabrication, the underlying hardware in various information systems that were once trusted can no longer be so. The untrusted chip fabrication and test facilities represent security threats to the current horizontal integration. The security threats posed by these entities include: ($i$) overproduction of ICs, where an untrusted foundry fabricates more chips without the consent of the SoC design house to generate revenue by selling them in the market~\cite{roy2008epic, alkabani2007active, chakraborty2008hardware, alkabani2007remote, huang2008ic, baumgarten2010preventing, GuinTODAES2016}, and ($ii$) piracy of IPs, where an entity in the supply chain can use, modify and/or sell functional IPs illegally~\cite{castillo2007ipp, tehranipoor2011introduction, Guin2015counterfeit, bhunia2018hardware}. An untrusted foundry has access to all the mask information constructed from the GDSII or OASIS files and then reconstructs all the layers and the complete netlist with advanced tools~\cite{torrance2009state}. In addition, reverse engineering (RE) of ICs becomes feasible even for advanced technology nodes due to the advancement of the tools for the decapsulation of the ICs and imaging. RE is commonly used in the semiconductor industry to perform failure analysis, defect identification, and verify intellectual property (IP)  infringement~\cite{torrance2007reverse, torrance2011state}. Unfortunately, the same RE can be exploited by an adversary to reconstruct the gate-level netlist from a chip~\cite{quadir2016survey}. 

 \begin{figure}[t]
    \centering 
    \includegraphics[width=\columnwidth]{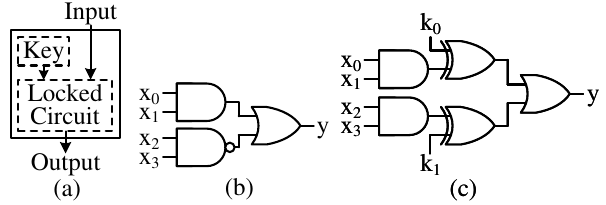} \vspace{-15px}
    \caption{\small Overview of logic locking. (a) Architecture of a locked circuit. (b) Original design. (c) XOR-based locking, with secret key of $k_0k_1=01$. 
    } \label{fig:logiclocking} 
     \vspace{-10px}
\end{figure}

One of the best ways to prevent an adversary from cloning a netlist (either by an untrusted foundry or a reverse engineer) is to hide or obfuscate the circuit. The attacker cannot decode the original functionality even after extracting the netlist from RE. Logic locking promises to hide the inner details of a circuit by inserting a set of key gates. The only way to recover the original functionality is by applying a secret key stored in a tamper-proof memory of the chip. Figure~\ref{fig:logiclocking} shows an abstract representation of logic locking. In addition to logic locking, hardware watermarking~\cite{charbon1998hierarchical, kahng2001constraint, qu2007intellectual} could identify and prevent copying a netlist to a certain extent; however, it does not offer a proactive protection mechanism. The initial efforts in logic locking~\cite{roy2008epic, baumgarten2010preventing, rajendran2012security,rajendran2015fault}, and hardware watermarking~\cite{charbon1998hierarchical, kahng2001constraint, qu2007intellectual} were broken by Boolean Satisfiability (SAT) attack~\cite{subramanyan2015evaluating}. The distinguishing input patterns, obtained from SAT solver, combined with their corresponding responses from the oracle, are crucial for SAT attack~\cite{subramanyan2015evaluating} to uniquely determine the secret key. A DIP with its oracle response is denoted as an input-output (IO) pair, and we will use this terminology throughout the paper. The effectiveness of SAT attack propels the research community for new locking schemes in the post-SAT era, which are summarized in Table~\ref{table:logiclocking}. These include point function-based lockings~\cite{yasin2016sarlock, xie2019anti, shamsi2018approximation, yasin2017ttlock, yasin2017provably, sengupta2020truly, shakya2020cas, yasin2017lock, zhou2021generalized, liu2020strong, zhou2017humble}, cyclic-based~\cite{shamsi2017cyclic, roshanisefat2018srclock, rezaei2019cycsat, roshanisefat2020sat, rezaei2018cyclic, yang2021looplock}, LUT/routing-based~\cite{kamali2018lut, kamali2019full, kolhe2019security, chowdhury2021enhancing, shamsi2018cross, patnaik2020obfuscating, sweeney2020modeling, kamali2020interlock}, scan and finite-state machine (FSM)-based lockings~\cite{zhang2017dynamically, karmakar2018encrypt, karmakar2019efficient, potluri2020seql, kamali2020scramble, rahman2021security, koushanfar2017active, dofe2017novel, meade2017revisit, roshanisefat2020dfssd, li2021janus, li2022janus}, timing-based~\cite{xie2017delay, zhang2018timingcamouflage, alam2019toic, sweeney2020latch, azar2021data, rahman2022oclock}, and high-level synthesis (HLS)-based~\cite{pilato2018tao, pilato2021assure, zuzak2021resource, muttaki2021hlock, limaye2021fortifying, karfa2021host}. Concurrently, multiple attacks~\cite{yasin2017removal, shamsi2017appsat, xu2017novel, shen2017double, shen2018sat, sirone2019functional, zhang2019tga, jain2020atpg, jain2021taal, limaye2021fa, sengupta2021breaking, alrahis2021gnnunlock, han2021does, limaye2022valkyrie, duvalsaint2019characterization, shamsi2019impossibility, shamsi2021praise, zhong2022afia, zhou2017cycsat, shen2019besat, shamsi2019icysat, azar2020nngsat, kamali2020interlock, alrahis2021untangle, alrahis2019scansat, limaye2019robust, limaye2020dynunlock, el2017reverse, shamsi2019kc2, roshanisefat2021rane, azar2022warm, hu2021fun, saha2021oracall, azar2019smt, chakraborty2018timingsat, azar2019smt, han2021does, limaye2022valkyrie} against these logic locking techniques arise. In addition, various Machine Learning-based attacks~\cite{chakraborty2019surf, chakraborty2021sail, sisejkovic2021challenging, alrahis2021omla}, which are structural in nature and do not require the oracle, target the identification and recovery of keys that are obfuscated after synthesis process by commercial CAD tools.

\begin{table}[t]
\caption{Summary of post-SAT logic locking techniques and corresponding attacks.} \vspace{-10px}\label{table:logiclocking}
\begin{center}
\centering 
\begin{tabular}{M{2cm}|M{2cm}|M{3cm}}
\hline 

\textbf{Locking Type} & \textbf{Techniques} & \textbf{Attacks} \\ \hline \hline
Point function   & \cite{yasin2016sarlock, xie2019anti, shamsi2018approximation, yasin2017ttlock, yasin2017provably, sengupta2020truly, shakya2020cas, yasin2017lock, zhou2021generalized, liu2020strong, zhou2017humble}  & \cite{yasin2017removal, shamsi2017appsat, xu2017novel, shen2017double, shen2018sat, sirone2019functional, zhang2019tga, jain2020atpg, jain2021taal, limaye2021fa, sengupta2021breaking, alrahis2021gnnunlock, han2021does, limaye2022valkyrie, duvalsaint2019characterization, shamsi2019impossibility, shamsi2021praise, zhong2022afia}\\ \hline

Cyclic  & \cite{shamsi2017cyclic, roshanisefat2018srclock, rezaei2019cycsat, roshanisefat2020sat, rezaei2018cyclic, yang2021looplock}  & \cite{zhou2017cycsat, shen2019besat, shamsi2019icysat} \\ \hline

LUT   & \cite{kamali2018lut, kamali2019full, kolhe2019security, chowdhury2021enhancing, shamsi2018cross, patnaik2020obfuscating, sweeney2020modeling, kamali2020interlock}  & \cite{azar2020nngsat, kamali2020interlock, alrahis2021untangle} \\ \hline

Scan   & \cite{zhang2017dynamically, karmakar2018encrypt, karmakar2019efficient, potluri2020seql, kamali2020scramble, rahman2021security} & \cite{alrahis2019scansat, limaye2019robust, limaye2020dynunlock} \\ \hline

FSM   & \cite{koushanfar2017active, dofe2017novel, meade2017revisit, roshanisefat2020dfssd, li2021janus, li2022janus}  & \cite{el2017reverse, shamsi2019kc2, roshanisefat2021rane, azar2022warm, hu2021fun, saha2021oracall} \\ \hline

Timing & \cite{xie2017delay, zhang2018timingcamouflage, alam2019toic, sweeney2020latch, azar2021data, rahman2022oclock} & \cite{azar2019smt, chakraborty2018timingsat} \\ \hline 
HLS   & \cite{pilato2018tao, pilato2021assure, zuzak2021resource, muttaki2021hlock, limaye2021fortifying, karfa2021host} & \cite{azar2019smt, han2021does, limaye2022valkyrie} \\ \hline \hline
\end{tabular}
\end{center} \vspace{-20px}
\end{table}

This paper presents two novel aspects to analyze the iteration complexity of the oracle-guided SAT attack. First, we show a detailed analysis of the SAT attack based on the conjunctive normal form (CNF) clauses stored in the SAT solver. The SAT attack iteratively finds DIPs to eliminate an equivalent class of incorrect keys. We explore what the attack learned after finding a DIP at each iteration. We show that the SAT tool creates a relationship between different key bits by applying the DIP to the oracle and observing the correct response. Note that the expected goal for any logic locking technique is to achieve an exponential iteration complexity of key size so that an adversary cannot determine the correct key value within given time constraints. However, our analysis points to the linear growth of the required patterns or iterations rather than the desired exponential increase with keys. Using examples, we show how the attack uses a DIP to eliminate a class of equivalent keys to make the complexity linear. We also show that the complexity gets even lower for circuits with multiple overlapping logic cones. Note that a logic cone can be described as a directed graph where the input nodes and gates point toward the sole output. Second, one interesting observation is that the complexity (\ie number of iterations/DIPs) of the SAT attack often reduces with increasing key size. We provide detailed explanations of why it takes fewer iterations to find the correct key when we lock a circuit with a larger key size. Finally, we analyze the SAT attack complexity for a circuit locked using point functions~\cite{yasin2017provably, xie2018anti, shakya2020cas}.

The contributions of this paper are summarized as follows:
\begin{itemize}

\item \textit{New perspective on SAT attack efficiency:} Even though the SAT attack was presented in 2015, its complexity analysis was not performed to find out why a DIP eliminates a large number of keys. This paper uses examples to describe the step-by-step analysis of incorrect key elimination for each DIP. The inter-dependencies among key bits are clearly revealed with a DIP and the corresponding oracle's response. We further show that the attack requires less number of iterations when keys can be observed simultaneously at multiple primary outputs. 

\item \textit{SAT attack complexity:} The majority of locking schemes focus on an exponential complexity close to the entire keyspace for ensuring hardness against SAT attack. However, SAT attack has shown an overall linear trend upon key size. Furthermore, increasing the number of key gates does not necessarily correlate to more iterations of solving the correct key. Instead, it is common to observe a local decrease in iteration complexity. \textit{To the best of our knowledge, we are the first to show the local reduction in attack complexity with a larger key size.} We believe that the findings of this paper provide researchers with the necessary information to develop an SAT-resilient solution. To address the reduction in attack complexity with a larger key, we observe that the oracle's response of a DIP plays an important role in removing a large number of incorrect keys. For example, a response 0 at the OR gate effectively splits the logic cone into two subcones, where keys inside the subcone are dependent, but independent from the other subcone's. Such IO pairs exponentially reduce the attack complexity. Similarly, logic 1 at the AND gate has a similar effect in shrinking the keyspace.

\item \textit{SAT analysis on point functions-based locking:} Logic locking with point functions has demonstrated a strictly exponential iteration complexity against SAT attack. Unfortunately, those locking designs with complementary key blocks can be broken under the proper constraining of sub-keys with the SAT tool. We show how and why SAT attack needs one IO pair only for deriving the complete key for AntiSAT under certain key restrictions, but it would remain exponential complexity if the constraint is placed on the other key block instead. We provide a similar analysis on CAS-Lock, which can effectively reduce the exponential iteration complexity to linear. We present insights on how the same analysis can be applied to TTLock and various versions of SFLL.

\item \textit{SAT attack time complexity:} 
The SAT attack, or its variants, can be very effective in breaking secure logic locking that aims to achieve exponential iteration complexity. To build an SAT-resilient solution, we investigate the time complexity rather than the iteration count. We show that the attack spends most time for the c6288 multiplier benchmark on the last iteration of UNSAT so to confirm that no other DIPs exist. Note that the iteration count is still linear to key size (see Table~\ref{tab:c6288}). 
\end{itemize}

\vspace{-5px}
The rest of the paper is organized as follows. We introduce the background of SAT attack and various locking methods in Section~\ref{sec:background}. The inter-dependency between keys learned by SAT solver after each DIP is extensively explored in Section~\ref{sec:analysisSectioncnf}. The SAT attack complexity is further analyzed and explained in Section~\ref{sec:complexity}. Analysis of the point functions is shown in Section~\ref{sec:pointfunction}. The future directions are described in Section~\ref{sec:future}. Finally, we conclude the paper in Section~\ref{sec:conclusion}.

\vspace{-5px}
\section{Background}\label{sec:background}

\subsection{SAT attack on Logic Locking}\label{sec:satattack}
The entire series of attacks and the solutions thereafter originated from the SAT attack~\cite{subramanyan2015evaluating}. Subramanyan \el~\cite{subramanyan2015evaluating} exploit the idea of combinational equivalence checking with miter circuit and Boolean Satisfiability~\cite{kuehlmann2002robust} to attack logic locking schemes. This oracle-guided attack successfully derives the secret key of various logic locking techniques~\cite{roy2008epic, baumgarten2010preventing,rajendran2012security,rajendran2015fault,charbon1998hierarchical, kahng2001constraint, qu2007intellectual} within a short time frame. The SAT attack requires two circuits, the original circuit, $C_O(X,Y)$, and its locked version, $C(X,K,Y)$, where $X$, $Y$, and $K$ are the inputs, outputs, and key, respectively. The correct key $K_c$ restores the original circuit functionality so that its output response is always consistent with the original circuit (e.g., the oracle) under every possible input combination, $C(X,K_c,Y)=C_O(X,Y)$. An incorrect key programmed in the tamper-proof memory leads to output mismatch under one or more input vectors. The output discrepancy between an incorrect key and the correct one is shown on the miter circuit's output. The SAT attack derives the key through the following steps:

\setlength{\textfloatsep}{5pt}
\begin{algorithm}[!ht]
\SetKwInOut{Input}{Input}\SetKwInOut{Output}{Output}
\Input{~Unlocked circuit, oracle ($C_O(X,Y)$) and locked circuit ($C(X,K,Y)$)}
\Output{~Correct Key ($K_c$)}
\vspace{-5px}
\nonl \rule{.45\textwidth}{0.4pt}

\SetAlgoLined

$i \leftarrow 0$\;
$F\leftarrow []$\; 

\While{($\mathtt{true}$) }{
	$i \leftarrow i+1$ \;
 	
	$[X_i, K_i,r] = \mathtt{sat}[F \wedge (Y_{A_i}\neq Y_{B_i})]$\;
	\If{ ($r == \mathtt{false}$)}{$\mathtt{break}$\;}
    	$Y_i = \mathtt{sim\_eval}(X_i)$\;
   $F\leftarrow F\wedge C(X_i,K,Y_i)$\;

}
$K_c\leftarrow K_{i}$\;
return $K_c$ \;
\caption{SAT attack on logic locking~\cite{subramanyan2015evaluating}.} \label{alg:satattack}
\end{algorithm} 

\noindent $\bullet$ \textit{Finding the distinguishing input pattern (DIP) from the miter circuit:} It first constructs a miter with two copies of the locked circuit $A$ and $B$. Both circuits ($C(X,K_A,Y_A)$ and $C(X,K_B,Y_B)$ in CNF) share the same input $X$ except for the keys, $K_A$, $K_B$. Any output mismatch between the two locked circuits can be easily identified at the miter's output. In each round (\ie $i^{th}$), the tool finds the hypothesis key $K_i$, and reports a Boolean indicator $r$ depending on whether a satisfiable assignment for the miter exists or not, Algorithm~\ref{alg:satattack}, Line 5. If SAT is returned, the miter succeeded in amplifying the mismatched output, $r$ is $\mathtt{true}$, and the corresponding input pattern $X_i$ is also recorded. 

\vspace{3px}
\noindent $\bullet$ \textit{Deriving the correct key:} Upon obtaining a DIP $X_i$, SAT attack acquires the actual output $Y_i$ from oracle simulation, $C_O(X_i,Y_i)$, Line 9. Input $X_i$ and output response $Y_i$ are used in updating the CNF formula $F$, Line 10. The clauses in $F$ help narrow down the valid keyspace until it is left with only the correct key(s). If the UNSAT conclusion is generated, the differential output cannot be observed, $r$ is assigned to $\mathtt{false}$, and $X_i$ is empty (Lines 6-8) and the program ends. Note that the last iteration of SAT attack returns UNSAT as all incorrect keys are pruned from the keyspace.

The SAT attack repeats the above two steps, where it iteratively checks for satisfiable assignment of the miter circuit. If $r$ is $\mathtt{true}$ at the $i^{th}$ iteration, we know that incorrect keys still exist in the search space. When the miter circuit becomes UNSAT with the clauses in $F$, the Boolean variable $r$ becomes $\mathtt{false}$, indicating no differential output exists. This means no more incorrect keys can be found as no discrepancy can be produced. If multiple keys remain in the search space, it must be true that multiple solutions are valid since they all give the same output response. This holds for a few locking designs~\cite{xie2018anti,shakya2020cas} and certain locking scenarios, \eg chained XOR key gates, where the correct key is not unique. Returning any one of them can restore the original circuit functionality. If only one key is left, it must be the right one. Then, the attack exits the $\mathtt{while}$ loop, Lines 6-8, and extracts the last round's hypothesis key as the correct one, Line~12. The attack finishes by reporting the correct key to the console, Line~13.

Note that the original SAT attack program~\cite{subramanyan2015evaluating} includes two preload vectors (all zeros and all ones) at the initial setup, before invoking SAT solver with the miter circuit. The number of IO pairs $|P|$ used to derive the correct key is one more than the number of total iterations $TI$, $|P|=2+ (TI-1)=TI+1$. This is because the last iteration does not produce a DIP. It is clear that the IO pair count for determining the secret key of a locked circuit is in the same order as the iteration count, which only differs by a constant of 1. For better analyzing the iteration complexity of SAT attack on c6288 benchmark (see Section~\ref{subsec:c6288}, we modify the original program~\cite{subramanyan2015evaluating} by disabling both preload vectors, resulting in $|P|=TI-1$. 

\vspace{-5px}
\subsection{SAT Resistant Logic Locking Techniques and Attacks} 

As SAT attack~\cite{subramanyan2015evaluating} successfully breaks various logic locking techniques~\cite{roy2008epic, baumgarten2010preventing,rajendran2012security,rajendran2015fault,charbon1998hierarchical, kahng2001constraint, qu2007intellectual}, it propels the research community to explore new locking schemes~\cite{yasin2016sarlock, xie2019anti, shamsi2018approximation, yasin2017ttlock, yasin2017provably, sengupta2020truly, shakya2020cas, yasin2017lock, zhou2021generalized, liu2020strong} that utilize point functions for achieving the minimal output corruptibility. SARLock~\cite{yasin2016sarlock} only perturbs one input pattern's output for each incorrect key. AntiSAT~\cite{xie2018anti, liu2020strong, zhou2021generalized} and CAS-Lock~\cite{shakya2020cas} configure the point function with two complementary blocks $g$ and $\overline{g}$. SFLL~\cite{yasin2017lock,yasin2017provably, sengupta2020truly} flips the output for certain input patterns, where the correct key flips back the upset output and restores the original functionality.  Although these techniques guarantee exponential iterations in SAT attack, various attacks~\cite{yasin2017removal, shamsi2017appsat, xu2017novel, shen2017double, shen2018sat, sirone2019functional, zhang2019tga, jain2020atpg, jain2021taal, limaye2021fa, sengupta2021breaking, alrahis2021gnnunlock, han2021does, limaye2022valkyrie, duvalsaint2019characterization} have been proposed to exploit the designs' vulnerabilities, \eg from structural and functional perspectives, and restore the original circuit. Nevertheless, SAT attack is still the backbone for the oracle-guided attacks~\cite{xu2017novel, shen2017double, shamsi2017appsat, shen2018sat, sirone2019functional, limaye2021fa, sengupta2021breaking,shamsi2021praise, shamsi2019impossibility}.

\section{SAT Attack Analysis: Pruning of Incorrect Key with CNF Update}\label{sec:analysisSectioncnf}
This section presents a novel perspective of analyzing the SAT attack's effectiveness in breaking various locking schemes in deriving the secret key. We investigate the CNF clauses stored in the SAT solver and how it gets updated in every iteration with a DIP and its output response. The CNF consists of multiple clauses connected with AND ($\wedge$). One or more literals are joined by OR ($\vee$) inside each clause. We use literals, variables, and nodes interchangeably. 

\begin{figure}[ht] 
\centering \vspace{-10px}
    \includegraphics[width=.8\columnwidth]{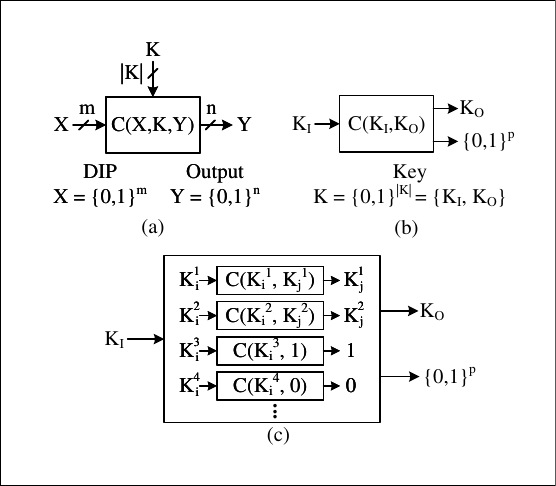}\vspace{-5px}
    \caption{\small Abstract representation of functions of key bits derived from an IO pair. (a) Locked circuit with an IO pair, (b) function of keys, and (c) subfunctions.}
    \label{fig:keyfunction} 
\end{figure}

The SAT attack requires an unlocked circuit, $C_O(X,Y)$, and its locked version, $C(X,K,Y)$, where $X$, $Y$, and $K$ are $m$, $n$ and $|K|$ bit wide. The correct key $K_c$ restores the original function so that its output response is always consistent with the unlocked circuit for all input combinations, \ie $C(X,K_c,Y)=C_O(X,Y)$. The SAT solver iteratively finds satisfiable assignments of the miter circuit whose inputs are denoted as DIPs. DIPs and the corresponding oracle outputs are denoted as IO pairs. As logic values for an IO pair $\{X,Y\}$ are known, $C(X,K,Y)$, shown in Figure~\ref{fig:keyfunction}(a), is transformed into the functions of keys $C(K_I,K_O)$, shown in Figure~\ref{fig:keyfunction}(b), where $K_O$ can be derived from $K_I$. Further, $C(K_I,K_O)$ can be expanded further and is shown in Figure~\ref{fig:keyfunction}(c). Any key in $K_O$, \eg $K_{j}^t$, is dependent upon key bits $K_{i}^t$, \ie $K_{j}^t=f(K_{i}^t)$, where $K_{i}^t \subseteq  K_I$. In addition, the combination of some key bits, \eg $K_{i}^s$, $K_{i}^s \subseteq  K_I$, produces a deterministic output, \ie either logic 0 or 1. 

\vspace{-15px}
\begin{align*}
    &C(K_I,K_O) \iff  \\
    & \begin{cases}
     C(K_{i}^t,K_{j}^t), \text{where } K_{j}^t=f(K_{i}^t), K_{j}^t \not\in K_{i}^t; i,j,t=1,2, \ldots\\
     C(K_{i}^s,\{0,1\}), \text{where } \{0,1\}=f(K_{i}^s); i,s = 1,2, \ldots
     \end{cases}
\end{align*} 

This key-dependent function $C(K_I,K_O)$ reveals additional information on the interdependency between key bits, \eg $K_{j}^t=f(K_{i}^t)$ and $\{0,1\}=f(K_{i}^s)$, crucial to the implicit removal of large incorrect key combinations. 

The placement of the key gates inside a particular cone is crucial as overlapping cones may reduce the attack complexity. A logic cone can be described as a combinational logic unit that represents a Boolean function bounded by an output and all its inputs. An increased number of primary outputs usually leads to multiple key values propagating across different output bits simultaneously. We begin our analysis with an example circuit with a non-overlapping cone with a single output and show how the SAT attack decrypts the 3-bit key with 3 IO pairs. Then, we describe how SAT attack can use fewer patterns to determine the secret key when key gates are placed under overlapping logic cones. 

\vspace{-10px}
\subsection{SAT Attack for a Locked Cone with One Output} \label{sec:cnf}
In this section, we examine how SAT attack implicitly removes the incorrect keys from the entire key search space. As described in Section~\ref{sec:satattack}, SAT solver finds a valid assignment to the miter circuit, and the tool records the extracted input vector, along with its output response obtained from the oracle simulation. The following example shows how SAT attack learns additional information on the secret keys from each IO pair from the miter circuit and oracle simulation. 

 \begin{figure}[th!]
    \centering 
    \includegraphics[width=\columnwidth]{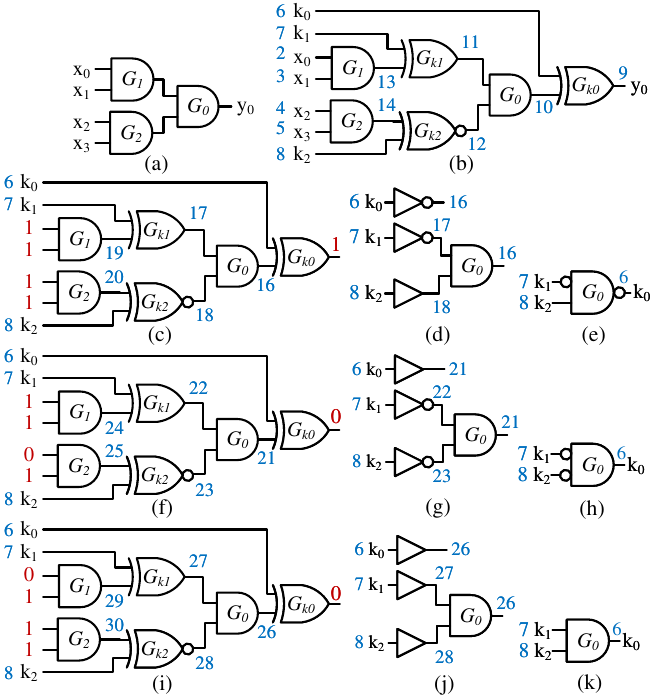}\vspace{-5px}
    \caption{\small Step-by-step SAT attack analysis. (a) Original circuit. (b) Locked circuit with $K=\{\text{001}\}$. CNF update and key-pruning for (c-e)  1$^{\text{st}}$ IO pair $P_1=\{1111;1\}$; 
    (f-h) 2$^{\text{nd}}$ IO pair $P_2=\{1101;0\}$, 
    and (i-k) 3$^{\text{rd}}$ IO pair $P_3=\{0111;0\}$.}
    \label{fig:singCNF} 
\end{figure}

Let us consider an example circuit with 4 inputs $x_0,...,x_3$ and 1 output $y_0$ of Figure~\ref{fig:singCNF}(a). Figure~\ref{fig:singCNF}(b) is the locked circuit with a 3-bit key, $k_0,k_1,k_2$, using strong logic locking (SLL) scheme. Each node is assigned a unique literal (in blue) by SAT solver. Upon finding a valid assignment to the miter circuit in the first iteration, DIP $X_1$ is extracted, $\{X_1\}=\{x_0,x_1,...,x_3\}=\{{1111}\}$. The output response $Y_1=1$ is obtained from oracle simulation with input $X_1$. SAT attack then records this IO pair $P_1=\{X_1;Y_1\}=\{x_0,...,x_3;y_0\}=\{{1111;1}\}$. We show in detail how the locked circuit's CNF gets updated under $P_1$, where the search space is shrunk in half (eliminated 4 incorrect keys). The literal assignment for the locked circuit's original CNF remains unchanged, but the internal nodes for IO pair $P_1$ are labeled with new variables \{16-20\} (Figure~\ref{fig:singCNF}(c), consistent with the internal operations of SAT attack~\cite{subramanyan2015evaluating}. The CNF for $C(X_1,K,Y_1)$ (abbreviated as $C_1$) is:
\vspace{-5px}
\begin{align*}
&C_1=
\overbrace{(\overline{17} \vee \overline{18} \vee 16 )\wedge 
(17 \vee \overline{16} )\wedge 
(18 \vee \overline{16} )}^{\text{AND gate }G_0}~\wedge \\ \nonumber&
\overbrace{(\overline{2} \vee \overline{3} \vee 19)\wedge 
(2 \vee \overline{19}  )\wedge 
( 3  \vee \overline{19})}^{\text{AND gate }G_1}~\wedge \\&
\overbrace{(\overline{4} \vee \overline{5} \vee 20  )\wedge 
(4 \vee \overline{20}  )\wedge 
(5 \vee \overline{20})}^{\text{AND gate }G_2}~\wedge \\ \nonumber & 
\overbrace{(\overline{6} \vee \overline{16} \vee \overline{9} ) 
\hspace{-2px}\wedge\hspace{-2px} (\overline{6} \vee 16 \vee 9)\hspace{-2px}\wedge\hspace{-2px}  
(6 \vee \overline{16} \vee 9)\hspace{-2px}\wedge \hspace{-2px}
(6 \vee 16 \vee \overline{9})}^{\text{XOR gate } G_{k0}} \wedge\\ \nonumber&
\overbrace{(\overline{7} \vee \overline{19} \vee \overline{17} )\hspace{-2px}\wedge \hspace{-2px}
(\overline{7} \vee 19 \vee 17)\hspace{-2px}\wedge \hspace{-2px}
(7 \vee \overline{19} \vee 17)\hspace{-2px}\wedge\hspace{-2px} 
(7 \vee 19 \vee \overline{17})\hspace{-2px}}^{\text{XOR gate } G_{k1}}\wedge \\\nonumber&
\overbrace{(\overline{20} \vee \overline{8} \vee {18}  )\wedge 
(\overline{20} \vee 8 \vee \overline{18} )\wedge 
(20 \vee \overline{8} \vee \overline{18}) \hspace{-2px}\wedge \hspace{-2px}
(20 \vee 8 \vee {18} )}^{\text{XNOR gate } G_{k2}}
\end{align*}
With IO pair $P_1$, we know the logic values for input/output, namely literals $2={1},3={1},4={1},5={1},9={1}$, and the $C_1$ is updated as:
\begin{eqnarray*}
C_1&\hspace{-10px}=\hspace{-10px}&
(\overline{17} \vee \overline{18} \vee 16 )\wedge 
(17 \vee \overline{16} )\wedge 
(18 \vee \overline{16} )\wedge 
(19)\wedge 
(20) \\ \nonumber&& \wedge~
(\overline{6} \vee \overline{16})  \wedge
(6 \vee 16)\wedge 
(\overline{7} \vee \overline{19} \vee \overline{17} )\hspace{-2px}\wedge\hspace{-2px}
(\overline{7} \vee 19 \vee 17)\\\nonumber&&\wedge~
(7 \vee \overline{19} \vee 17)\wedge 
(7 \vee 19 \vee \overline{17})\wedge
(\overline{20} \vee \overline{8} \vee {18}  ) \\\nonumber&&\wedge~
(\overline{20} \vee 8 \vee \overline{18} )\wedge 
(20 \vee \overline{8} \vee \overline{18} )\wedge 
(20 \vee 8 \vee {18} )
\end{eqnarray*}

Both nodes 19, 20 are in logic 1, and the CNF is adjusted:
\begin{eqnarray*}
C_1&\hspace{-10px}=\hspace{-10px}&
(\overline{17} \vee \overline{18} \vee 16 )\wedge 
(17 \vee \overline{16} )\wedge 
(18 \vee \overline{16} )\wedge 
(\overline{6} \vee \overline{16})\wedge  \\ \nonumber&& 
(6 \vee 16)\wedge 
(\overline{7} \vee \overline{17} )\wedge
(7 \vee 17)\wedge 
(\overline{8} \vee 18 ) \wedge
({8} \vee \overline{18}) 
\end{eqnarray*}\label{eqn:singCone}

\vspace{-10px}
This equation reveals that literals 6 and 16 have the opposite logic values, so are 7 and 17, while 8 and 18 are identical. The circuit representation of $C_1$ is shown in Figure~\ref{fig:singCNF}(d), still a function of $k_0,k_1$, $k_2$. These are the clauses appended in the formula $F$ (Algorithm~\ref{alg:satattack} Line 10). Equivalently, what SAT attack learned from the 1$^{\text{st}}$ IO pair $P_1$ is essentially a relation between 3 key bits, where $\overline{k_0}=\mathtt{AND}(\overline{k_1},k_2)$, as in Figure~\ref{fig:singCNF}(e). The constraint shrinks the possible keyspace in half.

On the second iteration, SAT attack returns the 2$^{\text{nd}}$ IO pair $P_2=\{X_2;Y_2\}=\{{1101;0}\}$, as in Figure~\ref{fig:singCNF}(d). Using the derivation we performed for the first iteration, the circuit representation of the added CNF clauses is shown in Figure~\ref{fig:singCNF}(g), which again is a function between the 3 key bits, $k_0=\mathtt{AND}(\overline{k_1},\overline{k_2})$ (Figure~\ref{fig:singCNF}(h)). It further shrinks the remaining keyspace in half, with only two keys left valid. Figure~\ref{fig:singCNF}(i) shows the 3$^{rd}$ IO pair $P_3=\{{0111;0}\}$ , whose CNF $C(X_3,K,Y_3)$ and its equivalent relation $k_0=\mathtt{AND}(k_1,{k_2})$ are illustrated in Figure~\ref{fig:singCNF}(j), (k), respectively. The combined effect of these three IO pairs, $P_1,P_2,P_3$, Figure~\ref{fig:singCNF}(e, h, k), uniquely determine key $K=\{k_0,k_1,k_2\}=\{{001}\}$. On the 4$^{\text{th}}$ iteration, no more distinguishing input can be found for the miter circuit, where $r$ is $\mathtt{false}$, and the SAT attack is complete.

In short, each IO pair provides additional information on the unknown key bits, where $C(X_i,K,Y_i)$ essentially becomes an equation for the unknown keys. A new equation for key is obtained in every iteration from the corresponding IO pair, which is independent of the findings derived from the previous rounds. SAT attack derives the secret key once the accumulated system of equations can uniquely determine all key bits.

\subsection{SAT Attack against Multiple Overlapping Logic Cones} \label{sec:multilogiccones} 
It is common for a circuit to have multiple outputs or fanouts. In other words, that circuit has multiple logic cones. With more fanouts, incorrect key responses are more likely to be observed than a single output. As the logic values for multiple keys can reach several outputs simultaneously, it accelerates and facilitates the removal of incorrect combinations to get the final key than the single logic cone where every key has to be observed from the same output pin. This is demonstrated by the example below. The following example shows that SAT attack needs fewer iterations to derive the secret key under multiple intersecting logic cones.

\begin{figure}[ht!]
    \centering 
    \includegraphics[width=\columnwidth]{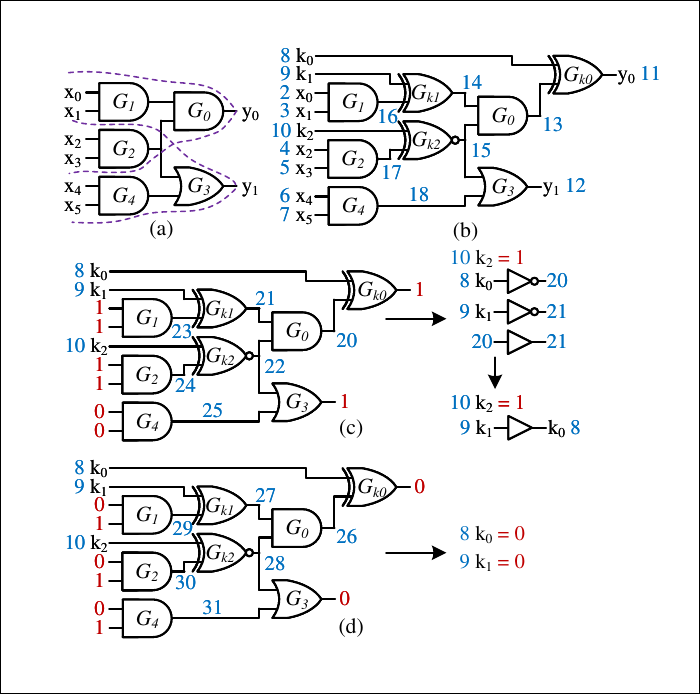}
    \caption{\small SAT attack on 2 intersecting cones with $K=\{001\}$. (a) Original circuit. (b) Locked circuit. CNF update and key-pruning for (c) 1$^{\text{st}}$ pair $P_1=\{{111100;11}\}$, (d) 2$^{\text{nd}}$ IO pair $P_2=\{{010101;00}\}$.} \label{fig:multiCNF} 
\end{figure}

Let us consider a circuit with 2 outputs, $y_0$ and $y_1$, as shown in Figure~\ref{fig:multiCNF}(a). The locked circuit, as shown in Figure~\ref{fig:multiCNF}(b), has 3 key bits, $k_0,k_1,k_2$, with the same locations as in Figure~\ref{fig:singCNF}(b). It differs from the locked circuit in Figure~\ref{fig:singCNF}(b) with additional gates $G_3$, $G_4$, and output $y_1$. This circuit has two logic cones; one with output $y_0$, inputs $x_0,x_1,x_2,x_3$, keys $k_0,k_1,k_2$, and gates $G_0,G_1,G_2,G_{k0},G_{k1},G_{k2}$; the other with output $y_1$, inputs $x_2,x_3,x_4,x_5$, key $k_2$, and gates $G_2,G_3,G_4,G_{k2}$. The effect of $k_2$ can be observed from both outputs, $y_0$ and $y_1$. SAT attack only needs 2 IO pairs to solve the keys, as opposed to 3 IO observations for the locked cone with a single output $y_0$ in Figure~\ref{fig:singCNF}(b). Figure~\ref{fig:multiCNF}(c) illustrates the 1$^{\text{st}}$ IO pair $P_1=\{X_1;Y_1\}=\{x_0,...,x_5;y_0,y_1\}=\{{011001;00}\}$, Its equivalent CNF expression of $C(X_1,K,Y_1)$ (abbreviated as $C_1$) is expressed in:

\vspace{-10px}
\begin{eqnarray*}
C_1\hspace{-10px}&=\hspace{-10px}&
(\overline{21} \vee \overline{22} \vee 20 )\wedge
(21 \vee \overline{20} )\wedge
(22 \vee \overline{20} )\wedge
(\overline{2} \vee \overline{3} \vee 23 )\\ \nonumber&&\wedge~
(2 \vee \overline{23} )\wedge
(3 \vee \overline{23} )\hspace{-2px}\wedge\hspace{-2px}
(\overline{4} \vee \overline{5} \vee 24 )\hspace{-2px}\wedge\hspace{-2px}
(4 \vee \overline{24} )\hspace{-2px}\wedge\hspace{-2px}
(5 \vee \overline{24} )\\ \nonumber&&\wedge~
(22 \vee 25 \vee \overline{12} )\wedge
(\overline{22} \vee 12)\wedge
(\overline{25} \vee 12)\hspace{-2px}\wedge\hspace{-2px}
(\overline{6} \vee \overline{7} \vee 25 )\\ \nonumber &&\wedge~
(6 \vee \overline{25} )\wedge
(7 \vee \overline{25} )\wedge
(\overline{8} \vee \overline{20} \vee \overline{11} ) \wedge
(\overline{8} \vee 20 \vee 11  )\\ \nonumber&&\wedge~
(8 \vee \overline{20} \vee 11  )\wedge
(8 \vee 20 \vee \overline{11}  )\wedge  
(\overline{9} \vee \overline{23} \vee \overline{21})\\ \nonumber
&&\wedge~
(\overline{9} \vee 23 \vee 21  )\wedge
(9 \vee \overline{23} \vee 21  )\wedge
(9 \vee 23 \vee \overline{21}  )  \\ \nonumber&&\wedge~
(\overline{10} \vee \overline{24} \vee {22})\wedge
(\overline{10} \vee 24 \vee \overline{22}  )\wedge
(10 \vee \overline{24} \vee \overline{22}  )  \\ \nonumber&&\wedge~
(10 \vee 24 \vee {22}  )  
\end{eqnarray*}

 \begin{figure*}[t]
    \centering 
    \includegraphics[width=\linewidth]{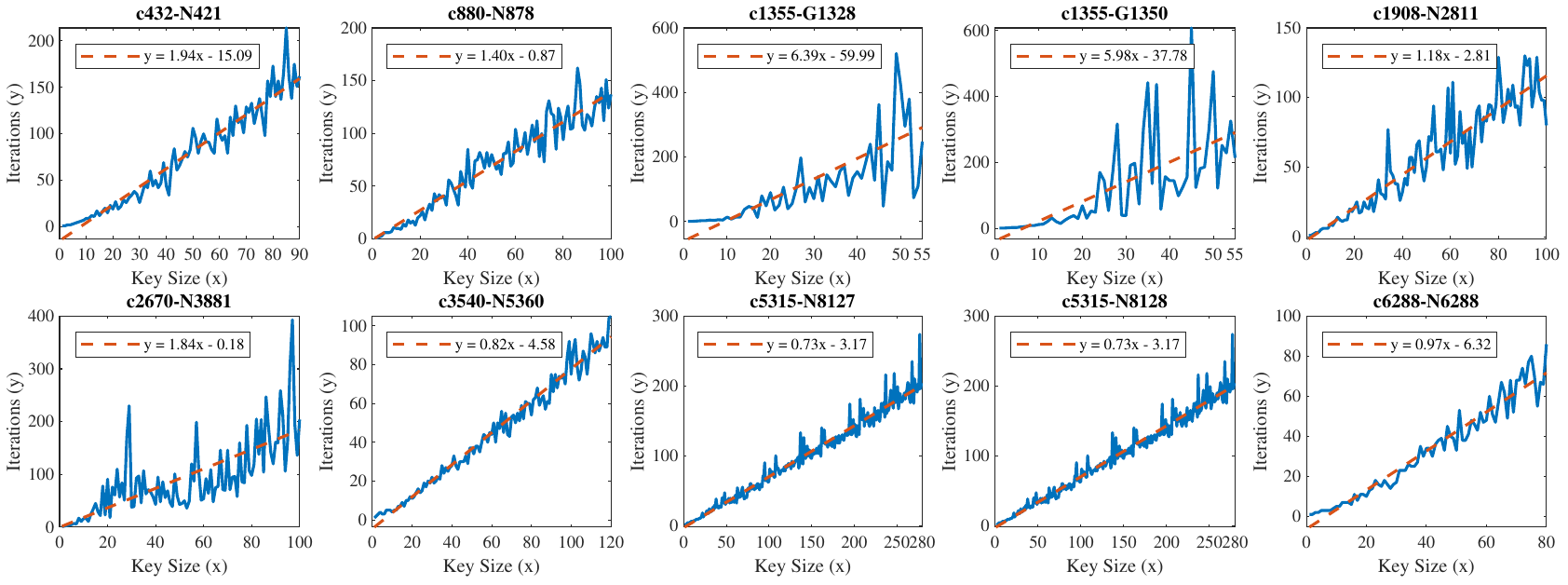}\vspace{-10px}
    \caption{\small SAT attack total iterations for ISCAS'85 benchmarks.} \label{fig:linIter} \vspace{-5px}
\end{figure*}

The 1$^{\text{st}}$ IO pair $P_1$ gives $2={1},3={1},4={1},5={1},6={0},7={0},15={1},17={1}$. The CNF for the locked circuit with $P_1$ is adjusted analogously to the previous example (Figure~\ref{fig:singCNF}) by plugging in the logic value of these known literals. It is straightforward that node 23, the output of AND gate $G_1$, has logic 1, as its inputs are literals $2={1}$ ($x_0$) and $3={1}$ ($x_1$). Similarly, nodes $24=1$, and $25=0$, based on literals $4-7$ ($x_2,...,x_5$). With an output 1 for OR gate $G_5$, its remaining input of node 22 must be 1, as $25=0$. Therefore, the CNF clauses added to SAT solver after the first iteration is:
\begin{eqnarray*}
C_1\hspace{-10px}&=\hspace{-10px}&
(\overline{21} \hspace{-3px}\vee\hspace{-3px} 20 )\hspace{-3px}\wedge\hspace{-3px}
(21 \hspace{-3px}\vee\hspace{-3px} \overline{20} )\hspace{-3px}\wedge\hspace{-3px}
(\overline{8} \hspace{-3px}\vee\hspace{-3px} \overline{20} ) \hspace{-3px}\wedge\hspace{-3px}
(8 \hspace{-3px}\vee\hspace{-3px} 20  )\hspace{-3px}\wedge\hspace{-3px}  
(\overline{9} \hspace{-3px}\vee\hspace{-3px} \overline{21})\hspace{-3px}\wedge\hspace{-3px}
(9 \hspace{-3px}\vee\hspace{-3px} 21  )\hspace{-3px}\wedge\hspace{-3px}
(10)
\end{eqnarray*}
With $C_1$, SAT attack determines key bit $k_2=1$, along with key-dependent equation $k_0=k_1$, as shown in Figure~\ref{fig:multiCNF}(c). With the 2$^{\text{nd}}$ IO pair $P_2=\{X_2;Y_2\}=\{{010101;00}\}$, as in Figure~\ref{fig:multiCNF}(d), SAT atttack uniquely determines both key bits $k_0$ and $k_1$ as logic 0. In the third round, SAT attack returns UNSAT (as all key bits are solved), and the program finishes. 

\begin{table}[ht!]
\caption{Comparison of SAT attack iterations ($TI$) between multiple primary outputs ($|PO|$) and single cone.} \label{tab:cktvscone}\vspace{-10px}
\begin{center}
\begin{tabular}{|c|c|c|c|c|c|c|}
\hline
\multirow{2}{*}{\textbf{Benchmark}} & \multicolumn{3}{c|}{\textbf{Locked Circuit (SLL)~\cite{subramanyan2015evaluating}}} & \multicolumn{3}{c|}{\textbf{Locked Cone (Sec.~\ref{sec:complexity})}} \\ \cline{2-7} 
      & {$\bm{|PO|}$}  & {$\bm{|K|}$} & $\bm{TI}$ & {$\bm{|PO|}$}  & {$\bm{|K|}$}   & $\bm{TI}$ \\ \hline
c432  &   7 & 80  &   24  &   1 & 21 &  27  \\ \hline 
c880  &  32 & 192 &   76  &   1 & 50 &  80       \\ \hline 
c1355 &  32 & 137 &   29  &   1 & 17 &  33      \\ \hline 
c1908 &  25 & 220 &   110 &   1 & 92 &  123      \\ \hline 
c3540 &  22 & 167 &   40  &   1 & 58 &  40       \\ \hline 
c5315 & 123 & 231 &   55  &   1 & 78 &  60       \\ \hline 
\end{tabular}
\end{center}
\end{table}

In addition to the example described above, we perform experiments to show a weaker attack resiliency for overlapping cones, as summarized in Table~\ref{tab:cktvscone} with locked ISCAS'85 benchmarks. Table~\ref{tab:cktvscone} compares the attack complexity with key sizes between the complete benchmark circuit, where multiple overlapping cones exist, and the extracted single cone from the same benchmark. Columns 2-4 and 5-7 list the number of primary outputs ($|PO|$), key size ($|K|$), and the total SAT attack iterations ($TI$) for breaking the SLL-based locked benchmarks and the corresponding largest cone, respectively. For example, SAT attack takes 76 iterations to determine a 192-bit key for the c880 benchmark, whereas it takes 80 iterations to break the largest cone of c880 locked with a merely 50-bit key. We observe the same behavior for all other benchmarks as well. The SAT attack can only break fewer keys (or smaller key size) for a single-output logic cone than for the keys of the same benchmark circuit having multiple PO. This confirms a lower complexity for overlapping cones, which is due to the effect of incorrect keys manifested through multiple outputs where the interdependency between key bits is broken. Therefore, having multiple overlapping logic cones will reduce the iteration counts for SAT attack, making it easier to derive the final key when key bits can be observed at the outputs simultaneously. Since we are examining and analyzing the effectiveness of SAT attack, we henceforth focus on the analysis with a single logic cone only, as it is more complex than multiple cones and offers an upper bound to the iteration complexity. If we show the linear iteration complexity for a non-overlapping cone, then automatically, the same linear complexity will be preserved for overlapping cones.

\begin{figure*}[t]
    \centering 
    \includegraphics[width=\linewidth]{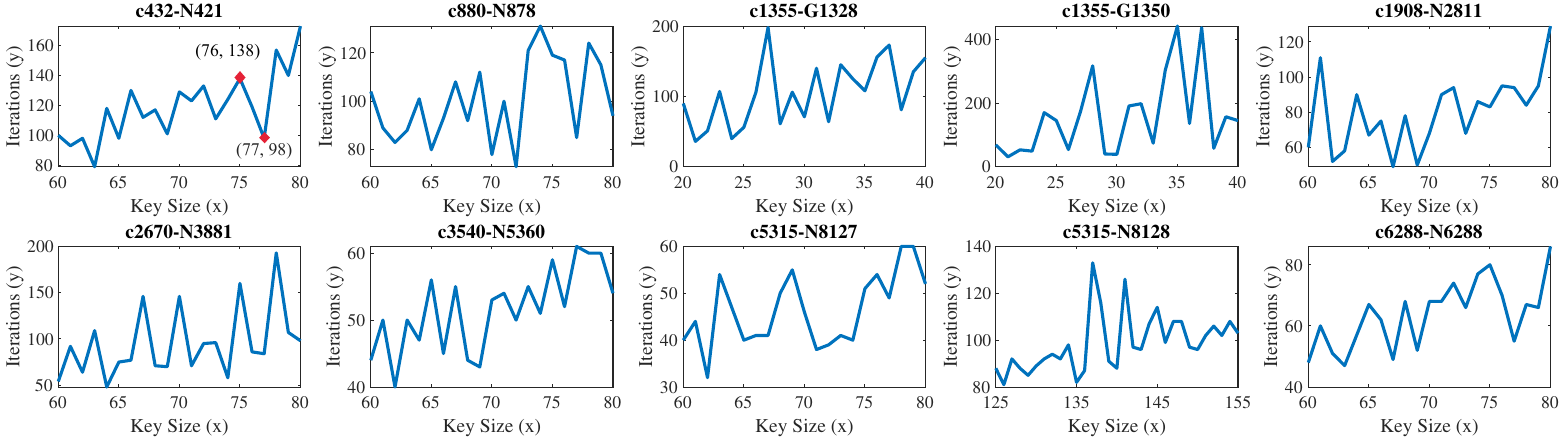}\vspace{-5px}
    \caption{\small The zoomed-in view of SAT attack Iterations for ISCAS'85 benchmarks.} \label{fig:zoomin} \vspace{-5px}
\end{figure*}

\begin{figure}[ht!]
    \centering 
    \includegraphics[width=\columnwidth]{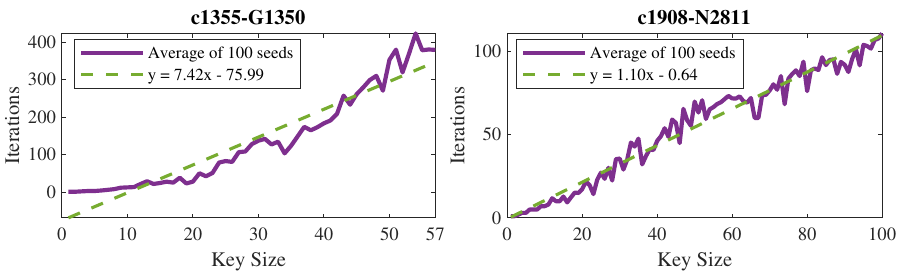} \vspace{-15px}
    \caption{The average iteration complexity of the SAT attack on locked benchmarks c1355-G1350 and c1908-N2811 with 100 different seed setups for SAT solver Lingeling.}
    \label{fig:avg-seed} 
    \vspace{-5px}
\end{figure}

\vspace{-5px}
\section{SAT Attack Analysis: Iteration Complexity}\label{sec:complexity}
In this section, we focus on the total iterations required for the SAT attack as the SAT attack complexity. We observe the linear iteration complexity for all ISCAS'85 benchmarks that agrees with the previously reported results. We, however, also observe the decrease in iteration complexity with increased key sizes for a large number of cases. To explain this phenomenon, we analyze how the output response from oracle, under certain DIPs, can trim more incorrect keys than other IO pairs. The complexity drop is caused by the multiple effective IO pairs selected by the tool. This explanation can also clarify the local peaks in iteration complexity due to the SAT attack selecting multiple less-effective IO pairs. We focus on the complexity trend for the iteratively increase in key sizes for any XOR-based locked circuits. The iterative insertion of keys (and key gates) ensures that the addition of one more key bit does not alter the locations of the already inserted key bits (and key gates). \textit{To the best of our knowledge, this is the first study to report the reduction of iteration count with increased key size.} 

The overview of SAT attack complexity analysis on the same logic cone is summarized in the following steps: ($i$) benchmark synthesis, ($ii$) cone analysis and the largest cone extraction, ($iii$) iterative insertion of key bits, and ($iv$) SAT attack iteration complexity aggregation. Synthesis is performed under 32nm technology libraries in Synopsys Design Compiler~\cite{SynopsysDC}. Figure~\ref{fig:linIter} shows an overall linear trend in total attack iterations under increased key sizes for non-overlapping cones. The best-fit lines are drawn in dashed lines with equations. To avoid the complexity reduction under multiple logic cones, the largest cone from each synthesized ISCAS'85 benchmark is extracted so that the response of any incorrect key combinations is observed through the sole output only. Each circuit is mapped to a directed graph with inputs pointing toward gates' output and, ultimately, the primary output. Logic cones are extracted by reversal of edge directions~\cite{flipedgeMatlab} and breadth-search~\cite{bfsearchMatlab} from each primary output. The ordered node list obtained in breadth-first search is used for determining ($i$) the largest cone (or cones if a tie) by node count and ($ii$) key gate insertion sequence as breadth-first search traverses all gates (nodes) within the same layer (same distance from output nodes) first before reaching gates at further layers. Following the same node order as in breadth-first search, we successively add one more XOR/XNOR key gate at a time, starting from gates closest to the primary output with increasing proximity. The original cone and its locked designs are all converted to the \textit{bench} format. SAT attack runs through all key sizes for every locked cone, and the total iterations are recorded. Figure~\ref{fig:linIter} shows the SAT attack iteration complexity on 9 benchmark cones with increasing key sizes. For example, c432-N421 is the logic cone from c432 benchmark with output N421. Cone c5315-N8127 and c5315-N8128 both contain the same gate count, but a significant overlap of gates exists. Please note that these logic cones all have reconvergent fanouts~\cite{juretus2019increasing}.

There are two observations from Figure~\ref{fig:linIter}. First, the overall complexity increase is not exponential, but linear. This means that, on average, the attack removes an exponential (or sub-exponential) number of incorrect keys per iteration. Second, all 9 benchmark cones exhibit the local non-monotonically complexity increase when additional keys are inserted. Note that a monotonic function ($f$) is either an entirely nonincreasing or nondecreasing function, where its first derivative does not change sign~\cite{royden1988real}. Now, $f$ is called monotonically increasing if $\forall x,y$, it satisfies $f(x) \leq f(y)$ for $x\leq y$. We denote a function as non-monotonically increasing if it increases globally (on average), but not monotonic. 

Figure~\ref{fig:zoomin} shows the zoomed-in view of SAT attack complexity for benchmark cones. For example, for cone c432-N421, it takes 138 iterations to break the key size of 76, but only needs 98 iterations when one more key bit is added. A non-monotonically increase in complexity is also observed in all the other benchmark cones. Note that the same non-monotonical behavior for the iteration complexity can still be observed under a different initialization seed setup. In addition, a non-monotonic linear increase can be observed in the averaged linear iteration complexity with 100 different seeds, as shown in Figure~\ref{fig:avg-seed} for c1355-G1350 and c1908-N2811. However, it does not suggest or infer that the minimum DIP count for solving each locked circuit with increased key sizes, a problem spans in PSPACE complexity instead. It is evident that the insertion of more key gates does not always lead to an increase in attack complexity. The non-monotonically increasing behavior in iteration complexity is observed in all cones. The question is, \textit{what causes the SAT attack to have such complexity drops when more keys are present in a locked design?} 

 \begin{figure}[t!]
    \centering 
    \includegraphics[width=\columnwidth]{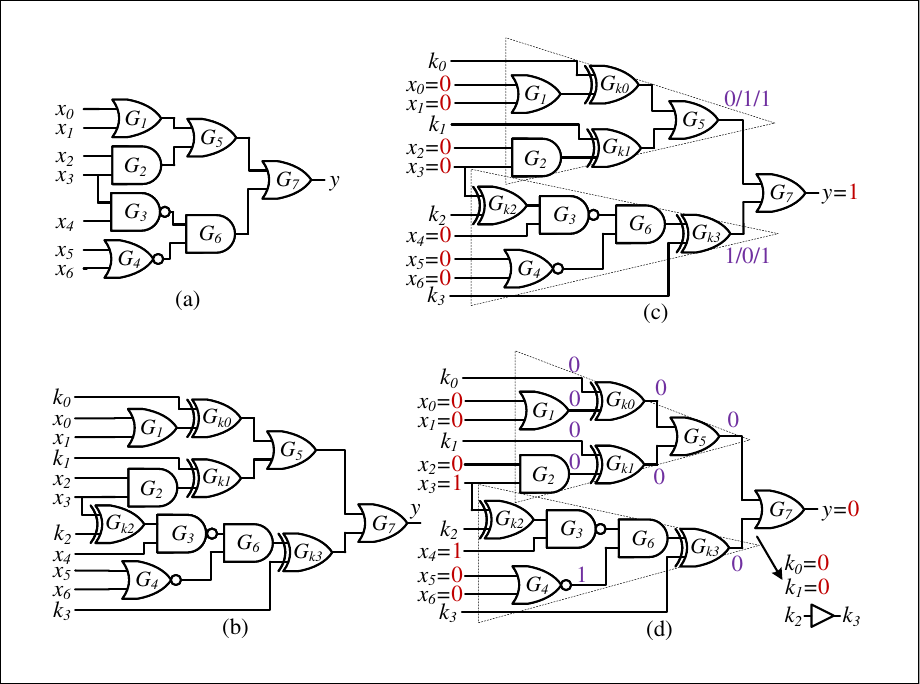}
    \caption{\small Key elimination. (a) original circuit, (b) locked circuit with 4 keys, (c) the $1^{st}$ IO pair $P_1=\{\text{0000000;1}\}$ from SAT attack (d) the second IO pair $P_2=\{\text{0001100;0}\}$ and equivalent relation of $k_0,k_1,k_2,k_3$ under $P_2$ only, where $k_0$ and $k_1$ are determined.} \label{fig:itc} 
\end{figure}

To describe the non-monotonically increasing behavior, we consider another example shown in Figure~\ref{fig:itc} where the effectiveness of individual IO pairs in eliminating incorrect keys is explored. The purpose here is to demonstrate that all the IO pairs are not equally effective in eliminating incorrect keys, some are better than others. As the SAT tool finds a DIP, which typically depends on the circuit topology, it is possible that the tool selects a more efficient DIP in earlier iterations for a locked circuit with a larger key that eliminates a large number of incorrect keys which results in a reduction in iteration. Figure~\ref{fig:itc}(a) shows the circuit, where four keys ($k_0-k_3$) are added (Figure~\ref{fig:itc}(b)). Note that an OR gate ($G_7$) is located at the cone output. The $1^{st}$ IO pair $P_1=\{X_1;Y_1\}=\{x_0,...,x_6;y\}=\{\text{0000000;1}\}$ returned by SAT attack has $y=1$ as the output. As logic 1 is the output of $G_7$, its two inputs could be any of the 3 combinations \{01/10/11\}. As the correct key cannot be determined uniquely, we focus on finding the incorrect ones, which are unique and result from \{00\}. One can simply find these incorrect ones using logic propagation, and it can be shown that there exist only 2 incorrect keys, Table~\ref{tab:truthtable} Column 2. Any key combinations that cause an output mismatch with the oracle's are marked with \xmark, indicating an incorrect key value implicitly removed from keyspace; key value(s) which produces the same output as the oracle's is noted with \cmark. From the 1$^{\text{st}}$ iteration, we observe fewer incorrect keys (i.e., $2\ll\frac{2^4}{2}$) are removed than the 2$^{\text{nd}}$ iteration (i.e., $14\gg\frac{2^4}{2}$) due to the properties of OR gate, where no unique conclusion can be made regarding its inputs (i.e., 10, 01 or 11) if the output is 1.

On the second iteration, the tool obtains another IO pair, $P_2=\{\text{0001100;0}\}$, with 0 at the output of the OR gate. The rest 13 incorrect key combinations are identified from $P_2$, as listed in Table~\ref{tab:truthtable}, Column 3. Here, we are interested in how $P_2$ trims more than half of the keys in the search space. The locked circuit with the IO pair $P_2$ is shown in Figure~\ref{fig:itc}(d). With the same derivation for CNF $C(X_2,K,Y_2)$, we know the outputs of gates $G_1,G_2,G_4$ are 0, 0, 1, once we have the input $X_2$. These gates' outputs can be similarly decided with $X_1$. With output $y=0$ at OR gate $G_7$, both inputs from this OR gate must be 0. This means both outputs of OR gate $G_5$ and XOR gate $G_{k3}$ are 0; and subsequently, the inputs of OR gate $G_5$ must be 0 as well, which are the output of both XOR gates $G_{k0}$ and $G_{k1}$. This results in the unique solution for 2 key bits $k_0$ and $k_1$ with $k_0=0,k_1=0$.

\begin{table}[t]
\caption{SAT attack uses 2 patterns to eliminate all 15 incorrect keys from keypace. If output differs from the oracle's, \xmark~is placed, else \cmark. The correct key is highlighted. \vspace{-10px}
}
\begin{center}
\begin{tabular}{ |c|c|c| } 
 \hline
 \textbf{4-bit key} & \textbf{IO Pair 1 ($\bm{P_1}$)} & \textbf{IO Pair 2 ($\bm{P_2}$)}\\
 $\bm{\{k_0,...,k_3\}}$ & \textbf{\{0000000;1\}} & \textbf{\{0001100;0\}}\\ \hline
 {\color{red}{\textbf{0000}}} & {\color{red}\cmark} & {\color{red}\cmark} \\ \hline
 0001 & \xmark & \xmark \\ \hline
 0010 & \cmark & \xmark \\ \hline
 0011 & \xmark & \cmark \\ \hline
 0100 & \cmark & \xmark \\ \hline
 0101 & \cmark & \xmark \\ \hline
 0110 & \cmark & \xmark \\ \hline
 0111 & \cmark & \xmark \\ \hline
 1000 & \cmark & \xmark \\ \hline
 1001 & \cmark & \xmark \\ \hline
 1010 & \cmark & \xmark \\ \hline
 1011 & \cmark & \xmark \\ \hline
 1100 & \cmark & \xmark \\ \hline
 1101 & \cmark & \xmark \\ \hline
 1110 & \cmark & \xmark \\ \hline
 1111 & \cmark & \xmark \\  \hline
\end{tabular}
\end{center}
\label{tab:truthtable}
\end{table}

A similar analysis can be performed on AND gates, whose inputs are uniquely defined under a logic 1 output. In summary, having a response of 0 at OR gates, or 1 at AND gates effectively splits the cone into two halves, where keys in one half are independent of the keys in the other half. This is equivalent to splitting the logic cone into two subcones based on input ports of OR/AND gates, where keys in both subcones can be evaluated and trimmed simultaneously. Therefore, the efficiency of removing incorrect keys depends on IO pairs, where the selection of an IO pair depends on the locked circuit topology that changes when adding more key bits. The effective IO pairs help remove more incorrect keys than the others. If a few effective IO pairs are selected in earlier iterations of the SAT attack, the iteration count can go down significantly. This leads to a non-monotonically increasing iteration complexity with the key size.   

\section{Case Study: Locking with Point Functions}\label{sec:pointfunction}
As the efficiency of SAT attack is indisputable, the subsequent logic locking proposals shift the focus toward building an exponential complexity in total iterations against SAT attack. One of the common approaches is to embed a point function right before the output of a logic cone, where the circuit's output response is perturbed based on the designer's chosen input combinations. This section presents a theoretical analysis of point functions of AntiSAT~\cite{xie2018anti}, CAS-Lock~\cite{shakya2020cas}, TTLock~\cite{yasin2017ttlock}, and SFLL~\cite{yasin2017provably}, and explains why they can also be broken by SAT-based attacks. In this section, we present a case study on how our proposed SAT attack analysis (see Sections~III, IV) can be used to analyze the attack complexity of KBM \& SAT~\cite{sengupta2021breaking}, a modified version of SAT attack with key constraints. The following analysis clarifies how and why SAT attack can still be effective in breaking AntiSAT and CAS-Lock under proper key constraints. Note that we are not proposing any new attacks but rather providing explanations to demonstrate that ($i$) AntiSAT with fixed $K_g$ requires only a single IO pair to determine the secret key, and ($ii$) the linear complexity for CAS-Lock under the same constraint. We also show that adding additional key constraints on $K_{\overline{g}}$ would not yield any extra benefits on the complexity reduction to an adversary for breaking AntiSAT-based locking designs. 
Note that the adversarial model for logic locking follows the same Kerckhoffs's principle as in modern cryptography, where the security of the system is based on the secret key and not on the obscurity of the algorithm used~\cite{alrahis2021gnnunlock, chakraborty2021sail, alrahis2021omla}. For point function-based locking techniques such as AntiSAT and CAS-Lock, we assume that the attacker has full knowledge of the locking scheme and the existence of a comparator logic inside the locked netlist. 

\subsection{Deterministic Property of SAT Attack with Constraints} \label{sec:pfexamples}
This section analyzes the SAT attack complexity on point function-based locking schemes with complementary blocks, $g$ and $\overline{g}$, where $K_{g}$ and $K_{\overline{g}}$ are inside $g$ and $\overline{g}$, respectively. Sengupta~\el~\cite{sengupta2021breaking} have shown an effective approach to reducing AntiSAT and CAS-Lock to polynomial complexity with key-bit mapping (KBM) \& SAT, where KBM separates $K_g$ and $K_{\overline{g}}$ and SAT attack is applied with a fixed $K_g$. The following analysis explains how and why SAT attack is still effective in breaking AntiSAT and CAS-Lock under proper key constraints. Our proposed SAT attack analysis also explains the same attack complexity of linear in iterations. In addition, we provide explanations to show ($i$) AntiSAT with fixed $K_g$ requires only one IO pair to determine the secret key, and ($ii$) linear complexity for CAS-Lock under the same constraint. However, constraining $K_{\overline{g}}$ would not give any extra benefits on the complexity reduction to an adversary on breaking AntiSAT.

\vspace{5px}
\subsubsection{SAT attack analysis on AntiSAT under key constraints} 
The two sets of keys, $K_{g}$ and $K_{\overline{g}}$, in AntiSAT offer two choices for the attacker, fixing one or the other. Using our key pruning analysis of Section~\ref{sec:analysisSectioncnf}, we explain how an adversary can determine the key with single IO pair only when setting $K_{g}$ constant. Yet, he/she will be less fortunate in breaking the secret key if $K_{\overline{g}}$ is kept constant instead.

\begin{figure}[t]
    \centering 
    \includegraphics[width=\columnwidth]{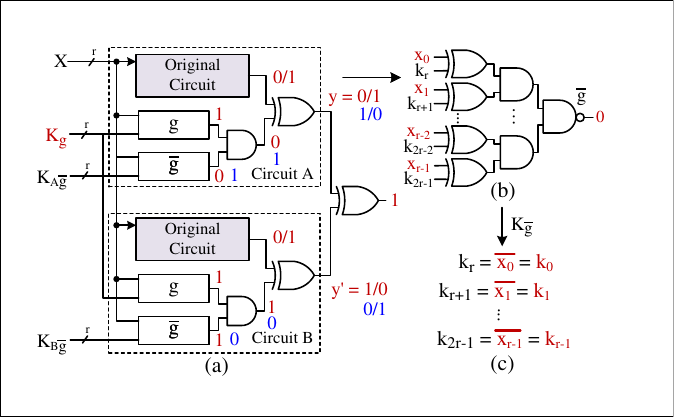} \vspace{-10px}
    \caption{\small SAT attack on AntiSAT with fixed key $K_g$. (a) Miter construction. (b) CNF update. (c)$K_{\overline{g}}=\{k_r-k_{2r-1}\}$ is determined.} \label{fig:constraints} 
\end{figure}

\vspace{5px}
\noindent $\bullet$ Key constraint on $K_{g}$: Let us consider a circuit with r-bit input $X=\{x_0,...,x_{r-1}\}$, 1-bit output $y$, locked with $2r$-bit keys of $K_{g}=\{k_0,...,k_{r-1}\}$ and $K_{\overline{g}}=\{k_r,...,k_{2r-1}\}$ of r-bit each. We assume the attacker already knows the bit locations for $K_g$ using the KBM of~\cite{sengupta2021breaking}. Furthermore, the r-bit $K_g=\{k_0,...,k_{r-1}\}$ is set to a constant vector. SAT attack is able to find an IO pair and uniquely determines all bits of $K_{\overline{g}}$. Figure~\ref{fig:constraints}(a) shows the miter construction, where $K_g$ are highlighted in red to indicate a fixed value. As miter creates differential output between two copies of the locked circuit A and B, without loss of generality, suppose the point function of circuit A has output 0 and circuit B's has output 1, shown in red (and vice versa in blue). Since both circuits have the same original cone, their output is identical to both A and B as they share the same input $X$. Without loss of generality, we assume the output of the original cone under the DIP found by the miter is logic 0. One can also assume with logic 1 instead. Hence, the output of AntiSAT block in circuit A is 0 while 1 for B's. As AntiSAT block has AND at the output, both inputs of this AND gate in B are 1, where $g=1$, $\overline{g}=1$. Then, we know that the DIP $X$ obtained from the miter must be complementary to the fixed $K_g$, $X=\overline{K_g}$ to ensure all ones for $g$'s AND tree of B. Following the analysis in Section~\ref{sec:analysisSectioncnf}, the solver updates its CNF clauses with $X$ and the oracle's output (logic 0 from the assumption). When this IO pair is applied to the locked circuit, the AntiSAT block gets a logic 0 output. Since DIP $X$ gives $g=1$ for circuit B, we still have $g=1$ during CNF update. Then, $\overline{g}=0$, as shown in Figure~\ref{fig:constraints}(b). As $\overline{g}$ is the NAND gate's output, all its inputs have logic 1. This uniquely determines all $r$-bit key $K_{\overline{g}}$, which is the complement of DIP $X$, $K_{\overline{g}}=X$, and identical to $K_g$, $K_{\overline{g}}=X=K_g$. SAT attack completes on the 2$^{\text{nd}}$ iteration since key $K_{\overline{g}}$ is already resolved. Therefore, the constraint on $K_g$ helps SAT attack finish within one IO pair. 

\noindent $\bullet$ Key constraint on $K_{\overline{g}}$: If the adversary decides to set key $K_{\overline{g}}$ constant instead, he/she will not get the same efficiency for key derivation as in fixing $K_g$. Suppose we constrain $K_{\overline{g}}=\{k_r,k_{r+1},...,k_{2r-1}\}$ to a constant r-bit vector. When SAT solver tries to find a satisfiable assignment to the miter circuit, following the same assumptions as before, we can derive that both A and B have the same logic 1 for $\overline{g}$ blocks. Having an output 1 at the NAND gate is equivalent to putting logic 0 to an AND gate. There are $2^r-1$ possible solutions for the r-bit input to produce a logic 1 at $\overline{g}$'s output. Equivalently, there are $2^r-1$ choices of DIP, satisfying the criterion of miter construction. When the tool updates SAT solver's CNF with DIP and output response, we get $\overline{g}=1$ for the NAND gate and $g=0$ for AND gate. Since unknown key bits are in $K_g$ of block $g$, a specific IO pair can prune only 1 incorrect key combination that results in $g=1(\neq0)$. The total IO pairs required to remove all incorrect keys of the r-bit keyspace for $K_g$ is $2^r-1$. The total iterations required for SAT attack is $2^r$. Therefore, by constraining $K_{\overline{g}}$, the adversary removes only one incorrect key and the overall SAT attack complexity remains exponential. 
 \begin{figure}[ht]
    \centering 
    \includegraphics[width=\linewidth]{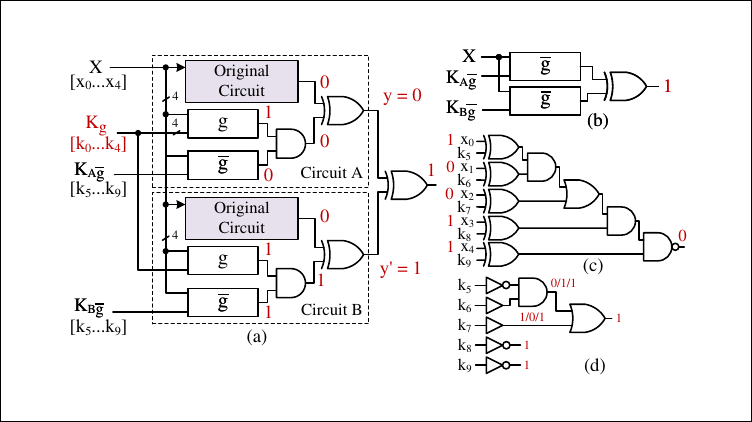}
    \caption{\small SAT attack on CAS-Lock with fixed $K_g$. (a) Miter construction and (b) Equivalent representation. (c) CNF update at SAT attack iteration 1. (d) Key pruning after iteration 1.} 
    \label{fig:caslock} 
\end{figure}

\subsubsection{SAT attack analysis on CAS-Lock under key constraints}
The same analysis on key constraints in AntiSAT can be applied to CAS-Lock, where the constraining of $K_g$ or $K_{\overline{g}}$ leads to linear complexity in solving $K_{\overline{g}}$ or $K_{g}$, respectively. We illustrate with a ($2r=10$)-bit CAS-Lock example, where block $g$ and $\overline{g}$ have one OR gate each, as shown in Figure~\ref{fig:caslock}. Our analysis can be generalized and applied to any OR gate replacement inside the cascaded AND chain of $g$ and $\overline{g}$. When SAT attack searches for a DIP $X$ for miter, as shown in Figure~\ref{fig:caslock}(a), the CAS-Lock block of one copy (\ie A) has logic 0 while the other (\ie B) has logic 1. As $K_g$ is fixed, the miter is essentially solving a differential output for $K_{\overline{g}}$ (Figure~\ref{fig:caslock}(b)). Suppose B's CAS-Lock block is 1, then it has $g=1$ and $\overline{g}=1$. The DIP $X$ obtained by SAT solver must satisfy $g=1$ as $K_g$ is constant. The oracle response, identical to the analysis for AntiSAT, helps to determine a logic 0 for the CAS-Lock block, as no alteration of output logic occurred. The CNF update implicitly eliminates the wrong keys in $\overline{g}$ with $\overline{g}=0$ under $g=1$ and output 0 for the combined blocks. After the 1st iteration, $k_8,k_9$ are uniquely determined, which reduces the key space from $2^5$ to $2^3$ and is shown in Figure~\ref{fig:caslock}(d). Note that some of the keys $k_5,k_6,k_7$ will be determined in the same way in the 2nd iteration of the SAT attack. The attack will continue iterating until all key bits are uniquely determined.

 \vspace{-10px}
\subsection{Extending the point function analysis to TTLock and SFLL} 
\begin{figure}[t]
    \centering 
    \includegraphics[width=0.70\columnwidth]{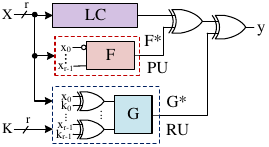}
    \caption{\small Generalized architecture of stripped functionality logic locking (SFLL). The functions F and G can be configured to implement TTLock~\cite{yasin2017ttlock} and SFLL-HD$^h$~\cite{yasin2017provably}. }
    \label{fig:sfll} 
\end{figure}

TTLock~\cite{yasin2017ttlock} and SFLL~\cite{yasin2017provably,sengupta2020truly} do not have two sets of keys like AntiSAT and CAS-Lock. Both perturb unit (\textit{PU}) and restore unit (\textit{RU}) are serially XORed with the original circuit, e.g., a logic cone (\textit{LC}) of interest, as shown in Figure~\ref{fig:sfll}. Keys are in the restore unit (\textit{RU}) only, where the perturb function ($F^*$) is key-free. 
The same analysis can be performed as PU and RU with the correct key implementing the same function even though different versions of SFLL have different output corruptibility. The output of PU, $F^*$, is logic 1 for only one input combination, where it alters the circuit behavior. The correct key helps flip back the perturbed logic and restores the original functionality as LC. Therefore, it must be true that the functional behavior for PU and RU are identical under the correct key so that LC's output is preserved. In other words, PU is the oracle for RU. If we can extract both PU and RU, we can then apply SAT attack on both circuits only, without requiring an oracle LC. As TTLock and SFLL perform logic synthesis after insertions of PU and RU, the adversary needs an accurate identification of PU and RU under logic optimization. The extraction of RU during post-synthesis is straightforward because commercial CAD tools cannot merge it inside LC or PU when the key is unknown; however, the challenging part is to retrieve PU since CAD tools may partially merge PU inside LC. Using the directed acyclic graph analysis~\cite{sirone2019functional}, there are multiple candidates for PU with full input $X$. One only needs to apply SAT attack to all possible PUs with the extracted RU and perform key validation in the end. \textit{Note that we do not need an unlocked chip (serves as the oracle for traditional SAT attack) as the oracle is already present in the synthesized LC \& PU circuit.} Our future work is to find an efficient way to determine the valid oracle and identify the wrong ones from all extracted PUs.

\section{Future Directions}\label{sec:future}

\subsection{Achieving Higher Time Complexity Against SAT Solvers}\label{subsec:c6288}
Even though point functions have demonstrated exponential iteration complexity, the adversary can formulate the attack with structural and functional analysis so that the complexity drops significantly. Although SAT attack has demonstrated linear trends in solving the secret key, we believe it is still possible for a logic design to achieve SAT resiliency. Here, we discuss how future locking schemes should consider a drastic increase in the hardness of their design against SAT attack from the example of c6288 benchmark. Besides targeting exponential iterations required for SAT attack, it may be feasible to significantly increase the overall time for SAT solver to find each satisfiable assignment for the miter circuit, which result in a longer computation time within each iteration.

\begin{table}[ht!]
\caption{Anatomy of SAT attack time on c6288\_N6288.} \vspace{-10px}
\begin{center}
\begin{tabular}{|M{.2cm}|M{.25cm}|r|r|r|r|M{1.1cm}|}
\hline
\multirow{2}{*}{\hspace{-4px}$\bm{|K|}$} & \multirow{2}{*}{\hspace{-4px}$\bm{|P|}$} & \multicolumn{4}{c|}{\textbf{CPU time (s)}} & \multirow{2}{*}{$\frac{\text{\textbf{UNSAT}}}{\text{\textbf{Total}}}\text{\textbf{(\%)}}$} \\\cline{3-6}
 &  & \multicolumn{1}{c|}{\textbf{Total}}& \textbf{IO Pairs}  &  \multicolumn{1}{c|}{\textbf{Average}} & \multicolumn{1}{c|}{\textbf{UNSAT}} & \\\hline
1  & 1  & 86.351 & 0.09108  & 0.09108  & 86.25948  & 99.894 \\ \hline
2  & 2  & 84.439 & 0.10289  & 0.05145 & 84.33634   & 99.878 \\ \hline
3  & 3  & 86.551 & 0.11019  & 0.03673 & 86.44092   & 99.872 \\ \hline
4  & 4  & 88.804 & 0.11963  & 0.02991 & 88.68458   & 99.865 \\ \hline
5  & 4  & 79.614 & 0.12705  & 0.03176 & 79.48717   & 99.840 \\ \hline
6  & 4  & 62.048 & 0.11630  & 0.02908 & 61.93153   & 99.812 \\ \hline
7  & 4  & 88.088 & 0.11822  & 0.02955 & 87.97006   & 99.865 \\ \hline
8  & 4  & 66.762 & 0.11330  & 0.02832 & 66.64887   & 99.830 \\ \hline
9  & 5  & 78.434 & 0.12385  & 0.02477 & 78.31049   & 99.842 \\ \hline
10 & 7  & 62.018 & 0.14788  & 0.02113 & 61.87004   & 99.761 \\ \hline
11 & 8  & 72.615 & 0.15925  & 0.01991 & 72.45534   & 99.780 \\ \hline
12 & 6  & 66.560 & 0.19532  & 0.03255 & 66.36468   & 99.706 \\ \hline
13 & 9  & 74.612 & 0.22130  & 0.02459 & 74.39026   & 99.703 \\ \hline
14 & 8  & 78.492 & 0.14760  & 0.01845 & 78.34455   & 99.811 \\ \hline
15 & 10 & 77.133 & 0.17205  & 0.01721 & 76.96051   & 99.776 \\ \hline
16 & 11 & 83.077 & 0.23765  & 0.02161 & 82.83926   & 99.713 \\ \hline
17 & 11 & 85.083 & 5.70418  & 0.51856 & 79.37841   & 93.295 \\ \hline
18 & 15 & 72.317 & 0.30082  & 0.02006 & 72.01650   & 99.584 \\ \hline
19 & 15 & 89.654 & 0.34619  & 0.02308 & 89.30831   & 99.613 \\ \hline
20 & 14 & 92.588 & 0.32586  & 0.02328 & 92.26268   & 99.648 \\ \hline
21 & 15 & 67.431 & 0.45250  & 0.03017 & 66.97835   & 99.328 \\ \hline
22 & 12 & 80.299 & 0.26642  & 0.02220 & 80.03259   & 99.668 \\ \hline
23 & 19 & 88.228 & 0.63912  & 0.03364 & 87.58904   & 99.275 \\ \hline
24 & 15 & 76.825 & 0.42104  & 0.02807 & 76.40369   & 99.451 \\ \hline
25 & 20 & 88.295 & 2.48402  & 0.12420 & 85.81125   & 97.186 \\ \hline
30 & 16 & 73.065 & 0.84954  & 0.05310 & 72.21507   & 98.837 \\ \hline
35 & 29 & 86.737 & 14.53748 & 0.50129 & 72.19920   & 83.239 \\ \hline
40 & 27 & 149.097 & 13.34636 & 0.49431 & 135.7502  & 91.048 \\ \hline
45 & 41 & 1130.466 & 18.31241 & 0.44664 & 1112.154 & 98.380  \\ \hline
50 & 37 & 84.404 & 6.16717 & 0.16668 & 78.23738    & 92.693 \\ \hline
55 & 45 & 1188.844 & 57.14645 & 1.26992 & 1131.698 & 95.193 \\  \hline
\end{tabular}
\end{center}\label{tab:c6288}
\end{table}

As described in SAT attack~\cite{subramanyan2015evaluating}, Subramanyan~\el stated that the multiplier benchmark c6288 is inherently challenging to SAT solvers, and was excluded from analysis. We locked its largest cone, N6288, in the same way as we did for other benchmarks in ISCAS'85, as described in Section~\ref{sec:complexity}. From the perspective of SAT attack iteration complexity, it remains linear with key size $|K|$, as shown on the bottom-right plot in Figure~\ref{fig:linIter} and Column 2 of Table~\ref{tab:c6288}. This suggests that c6288 behaves identically to other ISCAS'85 benchmarks. In addition, one can also observe that the complexity can decrease when more keys are inserted, as shown in Figure~\ref{fig:zoomin} and Column 2 of Table~\ref{tab:c6288}. The question is, \textit{what makes the circuit structure of a multiplier challenging to SAT solver?} To better analyze the SAT complexity in breaking c6288\_N6288, we record the CPU time spent for each iteration, including the very last UNSAT round. We exclude the pre-processing time, \ie setting up arrays of literals, initializing solver, \textit{etc}. The post-processing time is also excluded from the CPU time, \ie displaying the correct keys and the overall status, \textit{etc}. Table~\ref{tab:c6288} lists the time duration for SAT solver to derive the correct keys. The 1$^{\text{st}}$ and 2$^{\text{nd}}$ columns list key sizes $|K|$ and IO pair count $|P|$. Column 3 is the total time the SAT solver spent, which consists of two parts, ($i$) time used for generating all IO pairs, Column 4, and ($ii$) time checking that no more DIP exists (UNSAT), Column 6, where the averaged time it takes to find each IO pair is in Column 5. Column 7 reports the time ratio of the UNSAT decision over the total time spent on SAT solver. The interesting observation is that the major time spent was not on finding DIPs to prune off keyspace, but was on the last iteration, where SAT solver tries various backtracking before getting the UNSAT decision. The total time devoted to generating the IO pairs is negligible compared to the time spent in the very last iteration (UNSAT). In particular, the time duration for UNSAT in solving c6288 benchmark cones with respect to the total time span is generally over 90\%.

In summary, we believe it is possible to achieve SAT attack resiliency by using hard-to-find DIPs rather than SAT iteration count. We convey this message by presenting two different case studies of post-SAT locking with point functions. We showed the key pruning analysis on a modified version of the SAT attack, which employs the identification of key gates and their inputs, can eliminate an exponential number of incorrect key combinations with respect to total key space. In order to achieve SAT resiliency, one may need to incorporate the same SAT attack time complexity for each DIP as the last iteration of UNSAT in the c6288 multiplier benchmark. The objective is to considerably increase the total backtracks and logic reassignment required for SAT solvers to find a DIP in every iteration so that a longer time duration can be achieved. We conjecture that future locking schemes can provide sufficient difficulty for the present-day SAT solvers with conflict-driven clause learning (CDCL) algorithm~\cite{ganesh2020unreasonable}.

\vspace{-10px}
\subsection{Controllability Analysis} 
\begin{figure}[ht]
    \centering 
    \includegraphics[width=.95\columnwidth]{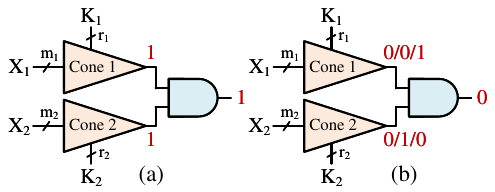} \vspace{-5px}
    \caption{SAT attack key evaluation of circuit with an AND gate at the output. Oracle response of (a) logic 1, (b) logic 0.}
    \label{fig:and01} 
\end{figure}

Controllability analysis can be incorporated prior to the key insertion to achieve a strictly monotonically increasing linear iteration complexity. Controllability, widely used in VLSI testing, is defined as the difficulty of assigning the target signal to a logic 0 or a logic 1~\cite{bushnell2004essentials}. A high value indicates the easiness of setting a node to that desired logic value from the inputs. Figure~\ref{fig:and01} shows an example of how controllability analysis can help analyze SAT complexity. If the output of AND gate is logic 1 with high probability (see Figure~\ref{fig:and01}(a)), its inputs can be uniquely determined, and the SAT attack can evaluate keys in parallel with the corresponding IO pair. This location is not a preferred location for inserting a key if the controllability of 1 is very high as many of the random input patterns will set this value. Instead, a very high probability of setting the same node to logic 0, as shown in Figure~\ref{fig:and01}(b), could be a desired location for placing the key gate.

Note that the defender's objective is to ensure DIPs are less effective in removing an exponential or sub-exponential number of incorrect keys (or finding a DIP). The goal is to keep the complexity in the order of $O(2^{K_1+K_2})$ (Figure~\ref{fig:and01}(b)), not $\mathtt{max}(O(2^{K_1}), O(2^{K_2}))$, equivalent to a logic 0 output at the AND gate in Figure~\ref{fig:and01}(a). We envision, with controllability analysis, nodes with different output probabilities for logic 0 and 1 under different gate types could be a good indicator for the adaptive key insertion strategy.

\vspace{-5px}
\subsection{Extension of the Proposed Complexity Analysis to the SMT Attacks}
Our complexity analysis can be extended to the SMT attacks proposed in~\cite{azar2019smt} due to the similarities between SMT and SAT. Satisfiability modulo theories (SMT), which consider the satisfiability of formulas under non-binary variables, offer more flexibility in the input space than the binary space for SAT. SMT attacks expand the capability of the SAT attack to target non-functional-based attacks such as delay and timing-based logic locking~\cite{xie2017delay}. Azar et al.~\cite{azar2019smt} proposed four approaches: ($i$) Reduced SAT Attack, ($ii$) Eager SMT Attack, ($iii$) Lazy SMT Attack, and ($iv$) Accelerated Lazy SMT Attack. Our future work will explore and extend the proposed complexity analysis approach to these SMT attacks. 

\vspace{-5px}
\section{Conclusion}\label{sec:conclusion}
In this paper, we provide a new perspective to analyze the efficiency of the SAT attack based on the CNF clause updates inside the SAT solver. In each iteration, SAT attack records the interdependencies between key bits from a distinguishing input pattern and its output response. Any locked circuit with multiple logic cones facilitates incorrect key removal as the effect of keys is propagated to multiple outputs. We further investigate the SAT attack complexity with the same cone of increasing key sizes. A non-monotonically increase in SAT complexity under increased key sizes is reported for the first time, where the insertion of additional key bits does not guarantee a strict linear growth in the SAT attack iteration complexity. Instead, this phenomenon of complexity drop happens to all ISCAS'85 benchmark cones. We subsequently provided an explanation of this observation from the oracle's response and logic gate types. It explains why more incorrect keys are eliminated from the keyspace with a particular IO pair. In addition, we give analytical reasoning to show how the constraining of key bits for post-SAT solutions like AntiSAT and CAS-Lock would aggressively reduce the key search down to constant or linear complexity. Finally, we furnish our discussions on SAT attack complexity analysis with novel observations on breaking the multiplier benchmark c6288, along with future directions.

\vspace{-5px}
\section*{ACKNOWLEDGEMENT}
This work was supported by the National Science Foundation under Grant Number CNS-1755733. 

\balance
\bibliographystyle{ieeetr} 

\vspace{-15px}
\begin{IEEEbiography}[{\includegraphics[width=1\linewidth]{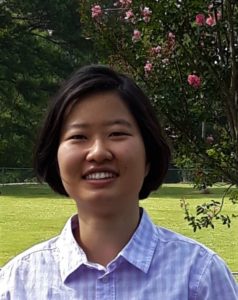}}]{Yadi Zhong (S'20)} is currently pursuing her Ph.D. in Computer Engineering from the Department of Electrical and Computer Engineering, Auburn University, AL, USA. She received her B.E. degree from the same university in 2020. Her research interests are logic locking, fault injection and hardware security, and post-quantum cryptography. She led a student team that received several awards including 1st place in HeLLO: CTF'21 and Hack@CHES 2021 and 2nd place in Hack@SEC 2021. She was the recipient of the Auburn University Presidential Graduate Research Fellowships in 2020. She is a student member of the IEEE.
\end{IEEEbiography}

\begin{IEEEbiography}[{\includegraphics[width=1in,height=1.25in,clip,keepaspectratio]{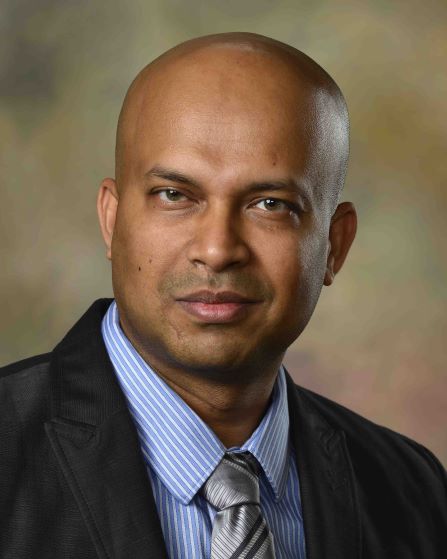}}]{Ujjwal Guin (S'10--M'16--SM'22)} received his PhD degree from the Electrical and Computer Engineering Department, University of Connecticut, in 2016. He is currently an Assistant Professor in the Electrical and Computer Engineering (ECE) Department of Auburn University, Auburn, AL, USA. He received his B.E. degree from the Department of Electronics and Telecommunication Engineering, Bengal Engineering and Science University, India, in 2004 and his M.S. degree from the ECE Department, Temple University, Philadelphia, PA, USA, in 2010. Dr. Guin's current research interests include hardware security, blockchain, and VLSI design \& test. He has authored several journals and refereed conference papers. He serves on organizing committees of HOST, VTS, ITC-India, and PAINE. He also serves on technical program committees in several reputed conferences, such as DAC, HOST, ITC, VTS, PAINE, VLSID, GLSVLSI, ISVLSI, and Blockchain. He is an IEEE senior member.
\end{IEEEbiography}

\end{document}